\documentclass[fleqn,usenatbib]{mnras}
\usepackage{newtxtext,newtxmath}
\usepackage[T1]{fontenc}
\usepackage{romannum}
\usepackage{graphicx} 
\usepackage[tight,footnotesize]{subfigure}
\usepackage{amsmath}	
\usepackage{multirow}
\usepackage{comment}
\usepackage{textcomp}
\hypersetup{
    colorlinks=true,
    linkcolor=blue,
    filecolor=blue,      
    urlcolor=blue,
    citecolor=blue,
}
\usepackage{afterpage}

\newcommand{\po}{P$_{\Omega}$\,} 
\newcommand{\ptwoo}{P$_{2\Omega}$\,} 
\newcommand{\pthreeo}{P$_{3\Omega}$\,} 
\newcommand{\pfouro}{P$_{4\Omega}$\,} 
\newcommand{\pfiveo}{P$_{5\Omega}$\,} 
\newcommand{\psixo}{P$_{6\Omega}$\,} 
\newcommand{\pseveno}{P$_{7\Omega}$\,} 
\newcommand{\peighto}{P$_{8\Omega}$\,} 

\newcommand{\rn}[1]{%
  \textup{\uppercase\expandafter{\romannumeral#1}}%
}
\title[An unusual eclipsing dwarf nova]{SRGA J115215.0−510656: an unusual long-period eclipsing dwarf nova with disc wind signatures}
\author[Rawat et al.]
{
Nikita Rawat $^{1}$\thanks{E-mail: rawatn@saao.ac.za, rawatnikita221@gmail.com},
David A.\,H. Buckley$^{1,2,3}$,
John R. Thorstensen$^{4}$,
Christian Knigge$^{5}$,
Yusuke Tampo$^{1,2}$,
\newauthor
Stephen B. Potter$^{1,6}$,
Anupam Bhardwaj$^{7}$,
Simone Scaringi$^{8}$,
Ilya A. Mereminskiy$^{9}$,
Jeewan C. Pandey$^{10}$,
\newauthor
Srinivas M Rao$^{10}$,
Alexander A. Lutovinov$^{9}$
\\
$^{1}$South African Astronomical Observatory, PO Box 9, Observatory, Cape Town, 7935, South Africa\\
$^{2}$Department of Astronomy, University of Cape Town, Private Bag X3, Rondebosch 7701, South Africa\\
$^{3}$Department of Physics, University of the Free State, PO Box 339, Bloemfontein 9300, South Africa\\
$^{4}$Department of Physics and Astronomy, 6127 Wilder Laboratory, Dartmouth College, Hanover, NH 03755-3528, USA\\
$^{5}$Department of Physics \& Astronomy, University of Southampton, Southampton SO17 1BJ, UK\\
$^6$Department of Physics, University of Johannesburg, PO Box 524, 2006 Auckland Park, Johannesburg, South Africa\\
$^{7}$Inter-University Centre for Astronomy and Astrophysics (IUCAA), Post Bag 4, Ganeshkhind, Pune 411 007, India\\
$^{8}$Centre for Extragalactic Astronomy, Department of Physics, Durham University, South Road, Durham DH1 3LE, UK\\
$^{9}$Space Research Institute, Russian Academy of Sciences, Profsoyuznaya str., 84/32, 117997, Moscow, Russia\\
$^{10}$Aryabhatta Research Institute of observational sciencES (ARIES), Nainital 263001, India\\
}

\date{Accepted XXX. Received YYY; in original form ZZZ}

\begin{document}
\label{firstpage}
\pagerange{\pageref{firstpage}--\pageref{lastpage}}
\maketitle

\begin{abstract}
We present the first detailed optical study of the cataclysmic variable SRGA~J115215.0$-$510656, based on new time-resolved photometric and spectroscopic observations complemented by long-baseline \textit{Transiting Exoplanet Survey Satellite (TESS)} data. The \textit{TESS} light curve reveals deep, recurring eclipses consistent with a high-inclination geometry and an orbital period of $0.43567659(9)$\,d. The eclipse morphology during outburst is consistent with a possible `inside-out' type outburst and supports classification of the system as a U~Gem--type dwarf nova.  By combining eclipse phase width and ellipsoidal modulation, we constrain the system geometry to a narrow locus in the $(q,i)$ plane, with allowed mass ratios $0.28 \lesssim q \lesssim 0.84$ and inclinations $i$ $\simeq$75--84$^{\circ}$. The persistence of single-peaked Balmer lines during outburst, together with strong He\,\rn{2}\,$4686$\,\AA\ emission and a flattened Balmer decrement, points towards emission arising in a disc wind or vertically extended regions above the disc. Absorption features from a late-type secondary star (approximately K3) are detected, contributing roughly 30 per cent of the red optical flux. Comparison with main-sequence expectations suggests that the donor star is moderately inflated, consistent with a mildly evolved secondary. With its long orbital period, modest outburst amplitude, and emission-line characteristics, SRGA~J115215.0$-$510656 appears to be a rare and compelling example of a bright, long-period dwarf nova whose optical properties are influenced by disc-wind processes during outburst.
\end{abstract}

\section{Introduction} \label{sec1}
Cataclysmic variables (CVs) are interacting binaries consisting of a white dwarf (WD) that accretes material from a Roche-lobe-filling companion (the donor), which usually resembles a late-type main-sequence star \citep[for a review, see][]{1995cvs..book.....W}. CVs are broadly classified as either magnetic or non-magnetic, depending on whether accretion is governed by the WD’s magnetic field. Among the non-magnetic systems, dwarf novae (DNe) are the most commonly observed subclass due to their recurrent outbursts \citep[for a review, see][]{1996osaki,2020hameury}. However, this is likely a selection effect, as DN outbursts constitute a primary discovery channel for such systems. During quiescence, material steadily accumulates in the accretion disc until a critical surface density is reached. At this point, a rapid increase in viscosity triggers an outburst, a behaviour described by the disc-instability model (DIM). In this model, outbursts are driven by a thermal instability developing in the accretion disc, while the mass-transfer rate from the donor is assumed to remain constant. DNe are further subdivided into three main classes -- SU UMa, Z Cam, and U Gem, based on their outburst properties, such as amplitude, duration, and recurrence timescale.  SU UMa-type systems typically lie below the traditional $\sim$2--3 h period gap of CVs. They display both normal outbursts ($\sim$2--5 days) and `superoutbursts', which persist for roughly 5--10 times longer. For these systems, however, an additional tidal instability is required, and the combined thermal–tidal disc-instability model \citep{osaki1989} provides a more complete framework for explaining their outburst behaviour. In contrast, Z Cam-type systems occasionally enter `standstill' phases, remaining $\sim$1 mag fainter than outburst maximum for several months.  
\par On the other hand, U~Gem-type systems are typically found above the period gap. They exhibit classical, quasi-periodic outbursts with amplitudes of 2--5~mag, lasting less than two weeks and recurring on widely varying timescales, most typically of order a few months \citep{2020hameury}. The prototype U~Gem exhibits two distinct types of outbursts -- `long' and `short', resulting in a bimodal distribution similar to that reported for other subtypes of DNe \citep{van-paradijs1983, cannizzo2002}. These outbursts typically last either $\sim$3--8~days (`short') or $\sim$12--16~days (`long'). One of the key reasons for the alternation between the two types of outbursts may be linked to fluctuations in the mass-transfer rate \citep{menard2001}, although similar behaviour can also be reproduced within the DIM framework \citep{mineshige1985}.
However, it is important to note that the outburst light curves of even the same system are not identical. Their shapes, durations, and rise rates depend on the radius at which the instability is triggered and on the type of outburst (inside-out or outside-in), and other system properties. In addition, DNe with smaller outburst amplitudes are generally observed to undergo outbursts more frequently \citep{Warner1987}.

\begin{figure*}
\centering
\subfigure[]{\includegraphics[width=0.9\textwidth,height=7cm]{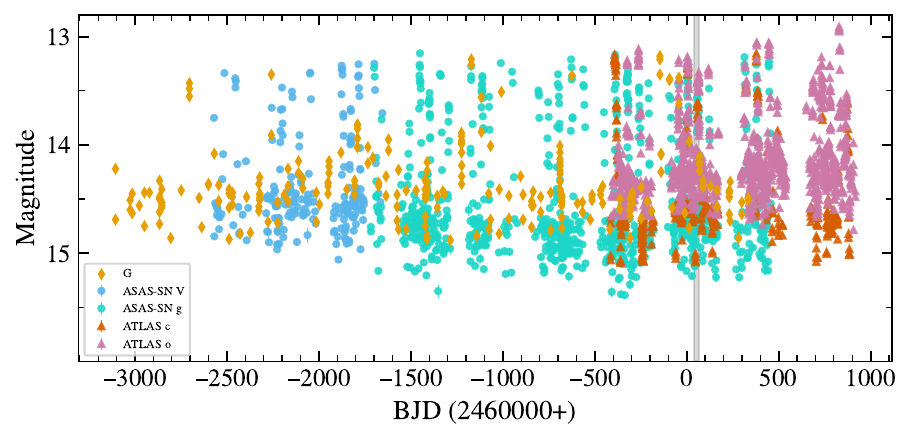} \label{fig:J1152_lc_all}}
\subfigure[]{\includegraphics[width=0.9\textwidth,height=10cm]{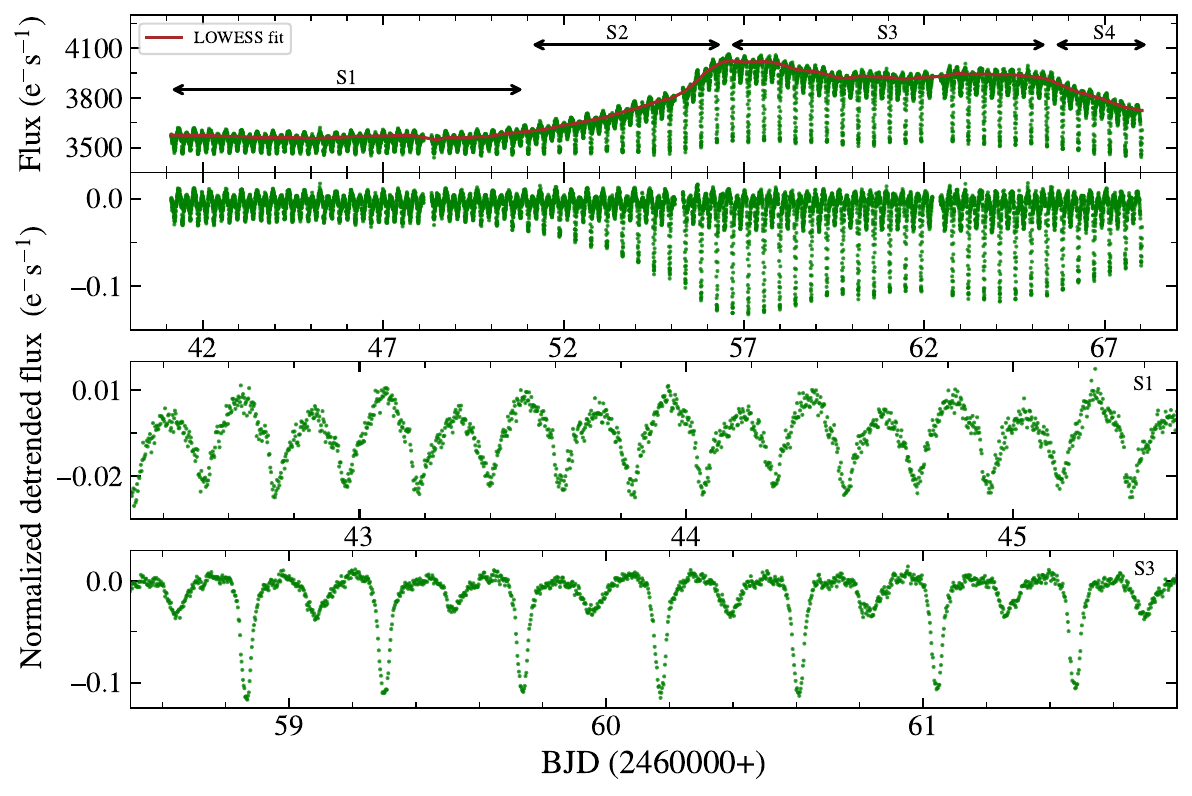} \label{fig:J1152_lc_tess}}
\caption{(a) Combined long-term \textit{Gaia}, ASAS-SN, and ATLAS light curve of J1152. The time span of the TESS observations is indicated in the grey shaded region. (b)  TESS light curve of J1152, where the solid brown line represents the smoothed light curve found using the LOWESS fit. The middle panel represents the detrended light curve after subtracting the smoothed light curve. The bottom panels show a zoomed-in view of the quiescent (S1) and outburst (S3) phases (see text for details).}
\label{fig:J1152_lc}
\end{figure*}

\par SRGA J115215.0−510656 was discovered in X-rays by the {\it Mikhail Pavlinsky} ART-XC telescope \citep{pavlinsky2021} onboard the Spectrum Roentgen Gamma (SRG) observatory \citep{2021A&A...656A.132S} during the course of its all-sky survey \citep{sazonov2024}. The source was classified as a candidate CV, based on the presence of known DN candidate ASASSN-V J115214.35−510700.5 (hereafter J1152) within the localization region \citep[see,][in preparation, for a CV selection routine]{Mereminskiy2025}. Adopting the \textit{Gaia} eDR3 parallax measurement, J1152 is estimated to lie at a distance of 640 $\pm$ 7 pc \citep{2021AJ....161..147B}. At this distance, its X-ray luminosity is about $L_{\mathrm{X}}$ $\approx$10$^{32}$ erg s$^{-1}$, typical for X-ray selected CVs. Its classification as a CV was confirmed through spectroscopic observations obtained with the South African Large Telescope \citep[SALT;][]{buckley2006, o'donoghue2006}.

\par In this paper, we present the first detailed investigation of the optical properties of J1152, making use of archival data from \textit{TESS} \citep{2015JATIS...1a4003R}, along with ground-based photometric and spectroscopic observations. This paper is organised as follows. In Section~\ref{sec2} we describe the observations and data reduction. Section~\ref{sec3} presents the analysis and results, followed by a discussion in Section~\ref{sec4}. Our conclusions are summarised in Section~\ref{sec5}.

\section{Observations and Data Reduction} \label{sec2}

\subsection{Spectroscopic observations} \label{sec2.3}
The identification spectrum of J1152 was obtained using the Robert Stobie Spectrograph \citep[RSS;][]{kobulnicky2003, burgh2003} mounted on SALT on 8 January 2025, with a total exposure time of 900~s, yielding a resolving power of R $\approx$600--900. Spectral extraction and wavelength calibration were performed using standard packages in IRAF. Relative flux calibration was achieved through observations of the spectrophotometric standard star EG~21, observed with identical grating settings and an exposure time of 60~s. In addition, time-resolved spectroscopic observations were conducted at the Sutherland Observatory of the South African Astronomical Observatory (SAAO) using the 1.9-m telescope equipped with a low-to medium-resolution spectrograph, known as SpUpNIC \citep{2019JATIS...5b4007C} during 13-24 February 2025. We used grating 6 (4400-7050 \AA) and a 1.34 arcsec slit, which yielded a spectral resolution near 4 \AA\ (FWHM; full-width at half-maximum). When conditions warranted, we observed flux standard stars from \citet{hamuy1992}.  We derived a master wavelength calibration from CuAr lamp spectra, and then used the night-sky lines at  5577 \AA ~and 6300 \AA, which were always present in the background, to derive a final first-order correction to the pixel-wavelength relation. We extracted 1D spectra using an independent Python implementation of the optimal extraction algorithm of \citet{horne1986b}\footnote{The extraction code is publicly available at \url{https://github.com/jrthorstensen/thorosmos}}, and flux-calibrated the data using standard IRAF tasks.

\subsection{Photometric observations} \label{sec2.1}
We observed J1152 simultaneously in the $B$ and $r$ bands using the 1.9-m and 1-m telescopes at the Sutherland Observatory of SAAO, with exposure times of 1 and 5~s, respectively. All exposures were GPS-triggered. The source was monitored over four nights between 13 and 16 June 2025 with an average coverage of 4.5~h per night. Data reduction was carried out in a standard manner using Python libraries, including {\tt photutils} v2.2.0 \citep{Bradley-2025} and the Astropy-affiliated package {\tt astroquery.gaia} \citep{Astropy-2013,Astropy-2018,Astropy-2022}. Calibrated magnitudes from aperture photometry were obtained using two field stars from the ESA Gaia Archive\footnote{\url{https://gea.esac.esa.int/archive/}}.

\subsection{\textit{TESS}, \textit{Gaia}, ASAS-SN, and ATLAS observations} \label{sec2.2}
J1152 was observed with \textit{TESS} in sector 64 from 2023 Apr 6 to May 4, at a cadence of 200 s in the Full Frame Images (FFIs). We obtained a 90$\times$90 pixel cutout centred on the target from the calibrated FFI using the TESSCut service\footnote{\url{https://mast.stsci.edu/tesscut/}}\citep{2019ascl.soft05007B}, at the Mikulski Archive for Space Telescopes (MAST). We performed aperture photometry with a 3$\times$3 pixel aperture centred on the target using the TESSreduce\footnote{\url{https://github.com/CheerfulUser/TESSreduce/tree/master}} package \citep{2021arXiv211115006R}.

\par We have also utilized the publicly available data of J1152 from \textit{Gaia}\footnote{\url{https://gsaweb.ast.cam.ac.uk/alerts/alert/Gaia19fnn/}} \citep{hodgkin2021}, the All-Sky Automated Survey for Supernovae \citep[ASAS-SN;\footnote{\url{https://asas-sn.osu.edu/variables}}][]{2014ApJ...788...48S, 2017PASP..129j4502K} and the Asteroid Terrestrial-impact Last Alert System \citep[ATLAS;][]{2018PASP..130f4505T, 2018AJ....156..241H, 2020PASP..132h5002S, 2021TNSAN...7....1S} to examine the long-term behaviour of the target.

\begin{figure}
\centering
\includegraphics[width=0.5\textwidth]{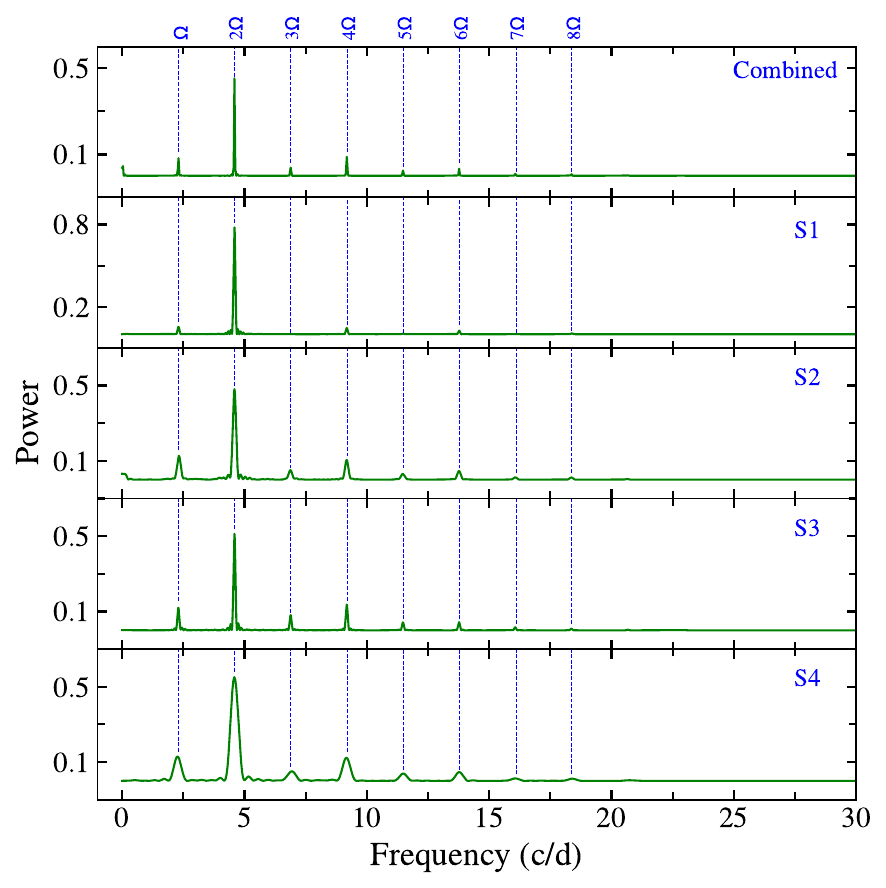}
\caption{LS periodogram of J1152 obtained from the combined detrended light curve from TESS. The bottom four panels represent the power spectra of different segments as shown in the top panel of Figure \ref{fig:J1152_lc}b (see text for details).}
\label{fig:tess_ps}
\end{figure}

\begin{table}
\centering
\caption{Eclipse mid-points for J1152 from observations from \textit{TESS} and SAAO.}
\label{tab:midpoint}
\renewcommand{\arraystretch}{1.4}
\begin{tabular}{lccc}
\hline
Eclipse mid-point (BJD) &  Cycle & Eclipse mid-point (BJD) &  Cycle \\
\hline
2460041.4428(42)    &    0  & 2460055.3803(11) &       32  \\   
2460041.8747(49) & 1 & 2460055.8151(9) &      33  \\ 
2460042.3139(46) & 2 & 2460056.2512(7) &      34 \\  
2460042.7514(51)   &     3  & 2460056.6868(6) &      35 \\    
2460043.1928(52) &        4 & 2460057.1226(7) &      36 \\    
2460043.6211(43) &       5  & 2460057.5575(7) &      37   \\  
2460044.0552(43) &       6  & 2460057.9934(7) &      38    \\ 
2460044.4920(46) &        7 & 2460058.4288(7) &       39  \\   
2460044.9299(45) &        8 & 2460058.8657(8) &      40   \\  
2460045.3671(46) &        9 & 2460059.3008(8) &       41    \\ 
2460045.8043(53) &       10 & 2460059.7367(8) &      42     \\ 
2460046.2343(39) &      11  & 2460060.1716(8) &      43      \\
2460046.6725(42) &       12 & 2460060.6072(8) &     44     \\ 
2460047.1064 (42) &      13 & 2460061.0434(9) &     45      \\ 
2460047.5420(40) &       14  & 2460061.4794(9) &      46  \\     
2460047.9813(48) &       15 & 2460061.9144(9) &       47 \\     
2460048.4121(43) &       16 & 2460062.7856(9) &       49 \\     
2460048.8531(50) &       17 & 2460063.2226(12) &     50   \\   
2460049.2852(49) &      18 & 2460063.6579(12) &     51    \\ 
2460049.7180(36) &       19 & 2460064.0932(13) &    52   \\   
2460050.1517(34) &       20 & 2460064.5287(12) &    53 \\    
2460050.5901(34) &      21 & 2460064.9643(12) &     54 \\   
2460051.0249(26) &      22 & 2460065.3993(12) &     55  \\   
2460051.4601(26) &       23 & 2460065.8360(13) &     56  \\     
2460051.8946(21) &     24  & 2460066.2712(14) &     57 \\   
2460052.3298(18) &       25 & 2460066.7072(14) &      58   \\    
2460052.7655(18) &      26  &  2460067.1428(14) &      59\\  
2460053.2018(17) &       27 & 2460067.5790(15) &      60\\      
2460053.6368(15) &       28 & 2460068.0143(22) &      61  \\  
2460054.0736(13) &       29  & 2460841.34031(3) &    1836\\    
2460054.5082(13) &       30 & 2460842.21194(5) &    1838  \\    
2460054.9441(12) &       31   \\       

\hline
\end{tabular}
\end{table}

\begin{figure*}
\centering
\includegraphics[width=0.9\textwidth]{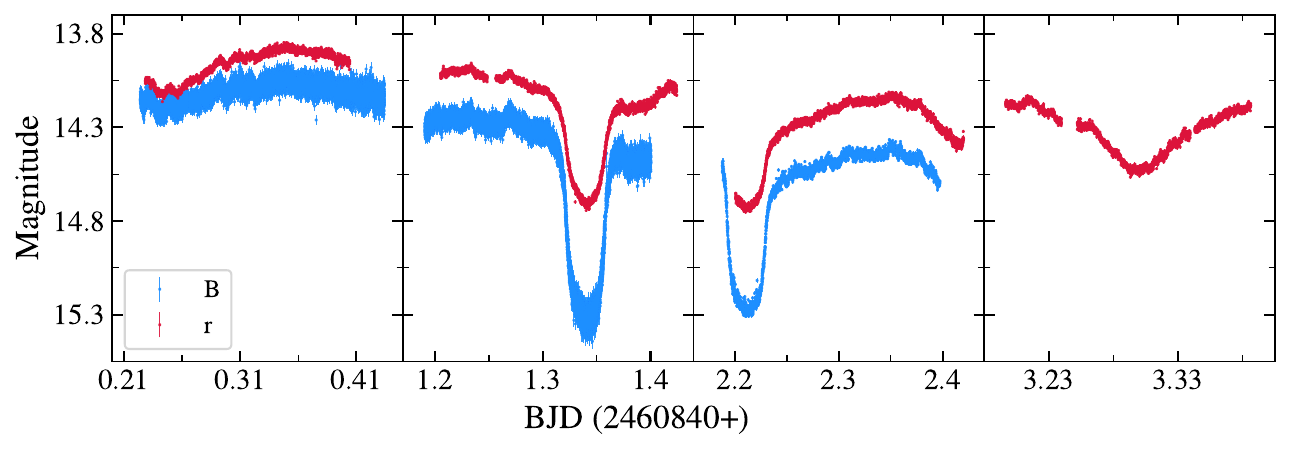}
\caption{Simultaneous B- and r-band light curves of J1152 obtained from SAAO during the quiescent phase.}
\label{fig:lc-saao}
\end{figure*}

\begin{figure*}
\centering
\includegraphics[width=0.9\textwidth]{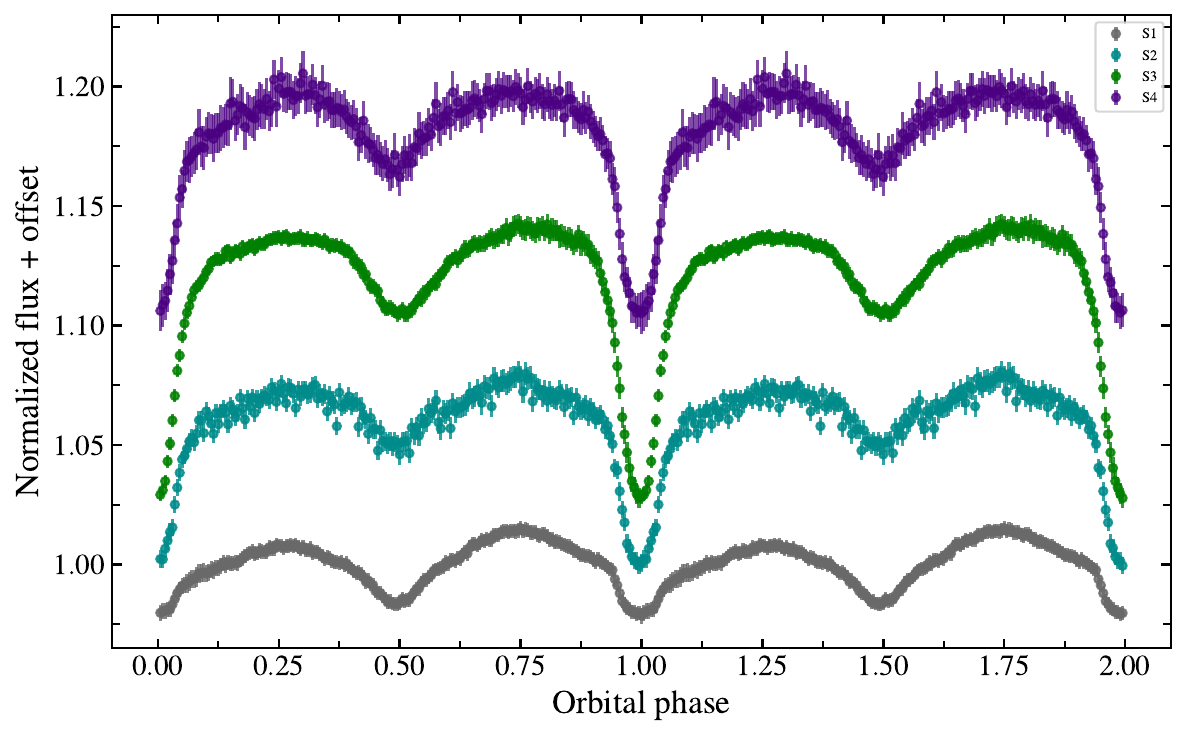}
\caption{Orbital phase-folded light curves of J1152 obtained from different segments of the TESS data, with phase bins of 0.005. For clarity, each successive segment is vertically offset by 0.06 in normalized flux.}
\label{fig:folded-tess}
\end{figure*}

\section{Analysis and Results} \label{sec3}

\subsection{Light curve morphology and power spectra} \label{sec3.1}
Fig. \ref{fig:J1152_lc_all} represents the long-term light curve of J1152 obtained using ASAS-SN V, ASAS-SN g, ATLAS c, and ATLAS o band data. The source exhibits clear outbursts, with an average outburst amplitude of approximately 1.6 mag and a recurrence timescale of roughly 40-60 days, indicating relatively frequent outbursts. The quiescent magnitude of J1152 is around 14.8 in the ASAS-SN V, ASAS-SN g, and ATLAS c bands, whereas it is approximately 14.5 in the ATLAS o band. The grey shaded region in Fig. \ref{fig:J1152_lc_all} indicates the time span of the \textit{TESS} observations.
\par The complete \textit{TESS} light curve of J1152 is shown in Fig. \ref{fig:J1152_lc_tess}, in which the source is observed transitioning from quiescence to an outburst state. We note that the \textit{TESS} flux is moderately affected by contamination from a nearby source (at a separation of $\sim$29$^{\prime\prime}$), at a level of $\sim$30 per cent, which may contribute to the smaller observed outburst amplitude compared to other surveys. However, this contamination does not affect the variability properties discussed in this work and there is no evidence that the nearby source is intrinsically variable. To remove the global trend, we first excluded the portions of the light curve corresponding to eclipses and then applied locally weighted polynomial regression \citep[LOWESS;][]{Cleveland1979} to obtain a smoothed curve, which was subsequently subtracted from the original data. The resulting detrended light curve is displayed in the middle panel of Fig. \ref{fig:J1152_lc_tess}, while the bottom panels provide a zoomed-in view of the quiescent and outburst phases. For detailed analysis, we divided the full \textit{TESS} light curve into four segments corresponding to the S1 (quiescence), S2 (rising), S3 (outburst), and S4 (declining) phases. The system remains in the outburst phase for approximately 7-8 days. As the system evolves from S1 to S3, the primary eclipses deepen significantly. To investigate periodic variability in J1152, we computed the power spectrum of the combined detrended light curve, as well as of the individual segments, using the Lomb-Scargle (LS) periodogram \citep{1976Ap&SS..39..447L, 1982ApJ...263..835S}, as implemented in the Astropy library. The resulting power spectra are shown in Fig. \ref{fig:tess_ps}, and the identified frequencies include $\Omega$, $2\Omega$, and higher-order harmonics up to the 8th. 
In each segment, the power spectrum is dominated by $2\Omega$ rather than $\Omega$, indicating a strong contribution from the donor star due to its ellipsoidal variations. 

\par We also observed J1152 on four occasions from 13-16 June 2025, during quiescence, and the resulting light curves are shown in Fig. \ref{fig:lc-saao}. Two primary eclipses were detected on 14 and 15 June, and one secondary eclipse on 16 June. The out-of-eclipse quiescent color was measured as B$-$r $\simeq$ 0.2 mag.

\subsection{Ephemeris and nature of eclipse profiles} \label{sec3.2}
The eclipse ephemeris of J1152 was derived from the timings of the eclipse minima. The mid-points of the eclipses were determined by fitting the observed light curves with a model composed of a constant baseline and a Gaussian function. In total, we obtained 61 eclipse timings from \textit{TESS} data and 2 additional minima from our own observations at SAAO. The newly derived times of minima are listed in  Table \ref{tab:midpoint}, with the associated uncertainties given in parentheses. All times are reported in the Barycentric Julian Date (BJD) reference frame. A linear fit between the cycle numbers and minima timings provided the following ephemeris for J1152:
\begin{equation} \label{eqn}
T_{0} = 2460041.4381(2) + 0.43567659(9) \times E   
\end{equation}
where $T_{0}$ is defined as the time of mid-eclipse and the errors are given in parentheses. The orbital period in equation (\ref{eqn}) is equivalent to 10.456238 $\pm$ 0.000002 h.

\par
\begin{figure}
\centering
\includegraphics[width=0.5\textwidth]{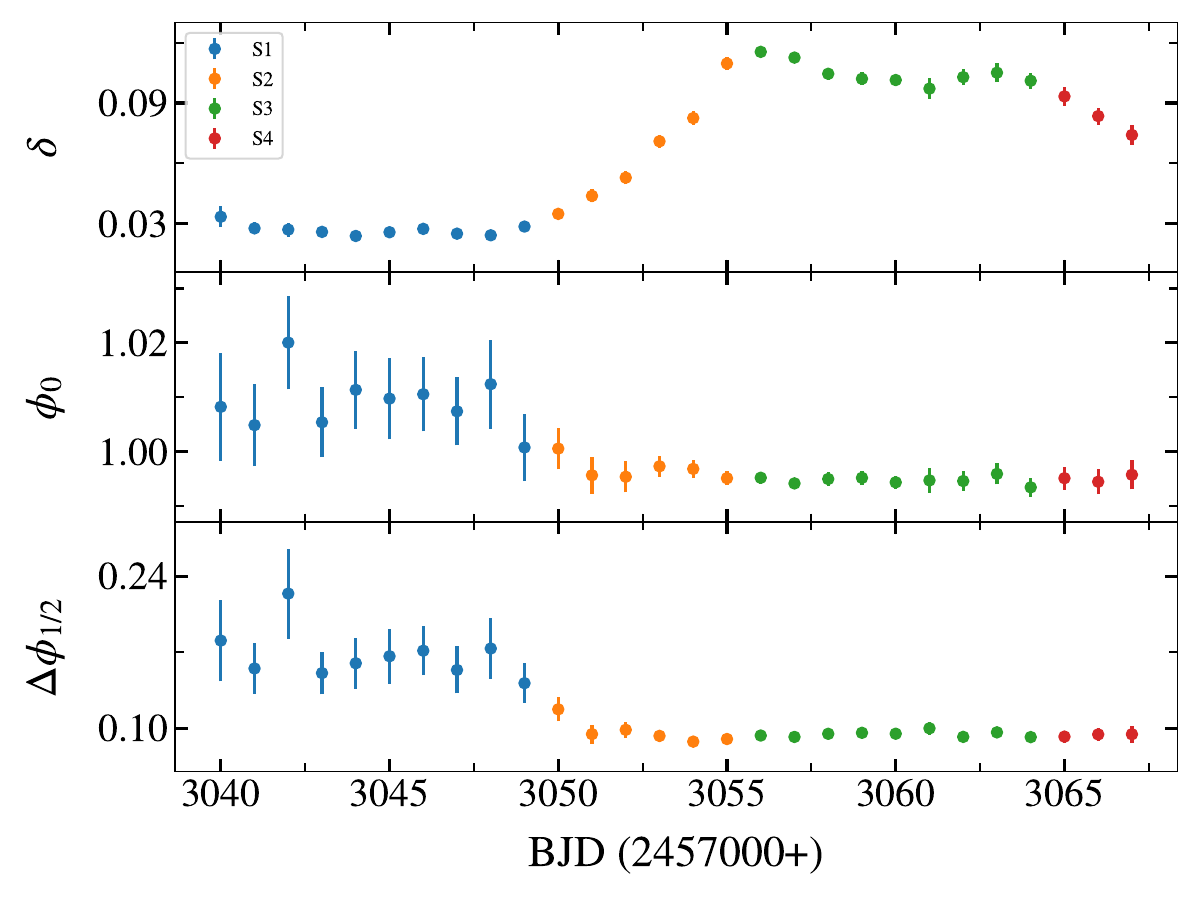}
\caption{Fractional eclipse depth, mid-eclipse phase, and eclipse phase width for the S1, S2, S3, and S4 segments of the \textit{TESS} light curve.}
\label{fig:eclipse}
\end{figure}

 To investigate the evolution of eclipse profiles across different segments of the \textit{TESS} data, we folded the light curves using the ephemeris given in equation~(\ref{eqn}). The orbital phase-folded light curves are presented in Fig. \ref{fig:folded-tess}, which clearly illustrate the variations in eclipse morphology. From S1 to S4, both the depth and shape of the eclipses exhibit significant changes. During S1, the system shows W~UMa-type orbital modulations together with pronounced hump features, indicative of a hotspot contribution. The primary eclipses in this state are shallow, U-shaped, asymmetric, and comparable in depth to the secondary eclipses. In contrast, during S2, the eclipse depth begins to increase and the profiles become more distinctly V-shaped. Further, during S3, the out-of-eclipse brightness remains nearly constant, while the eclipses appear symmetrical and distinctly V-shaped, consistent with the typical behavior of eclipsing DNe during outburst \citep[e.g.,][]{baptista2000b, kimura2018, joshi2024}. As the system transitions into S4, the eclipse depth decreases once again (see also the top panel of Fig. \ref{fig:eclipse}). These progressive changes in eclipse morphology most likely reflect structural variations in the accretion disc and in the hotspot contribution as the system evolves from quiescence through outburst and back.

\begin{figure}
\centering
\includegraphics[width=0.5\textwidth,height=0.5\textwidth]{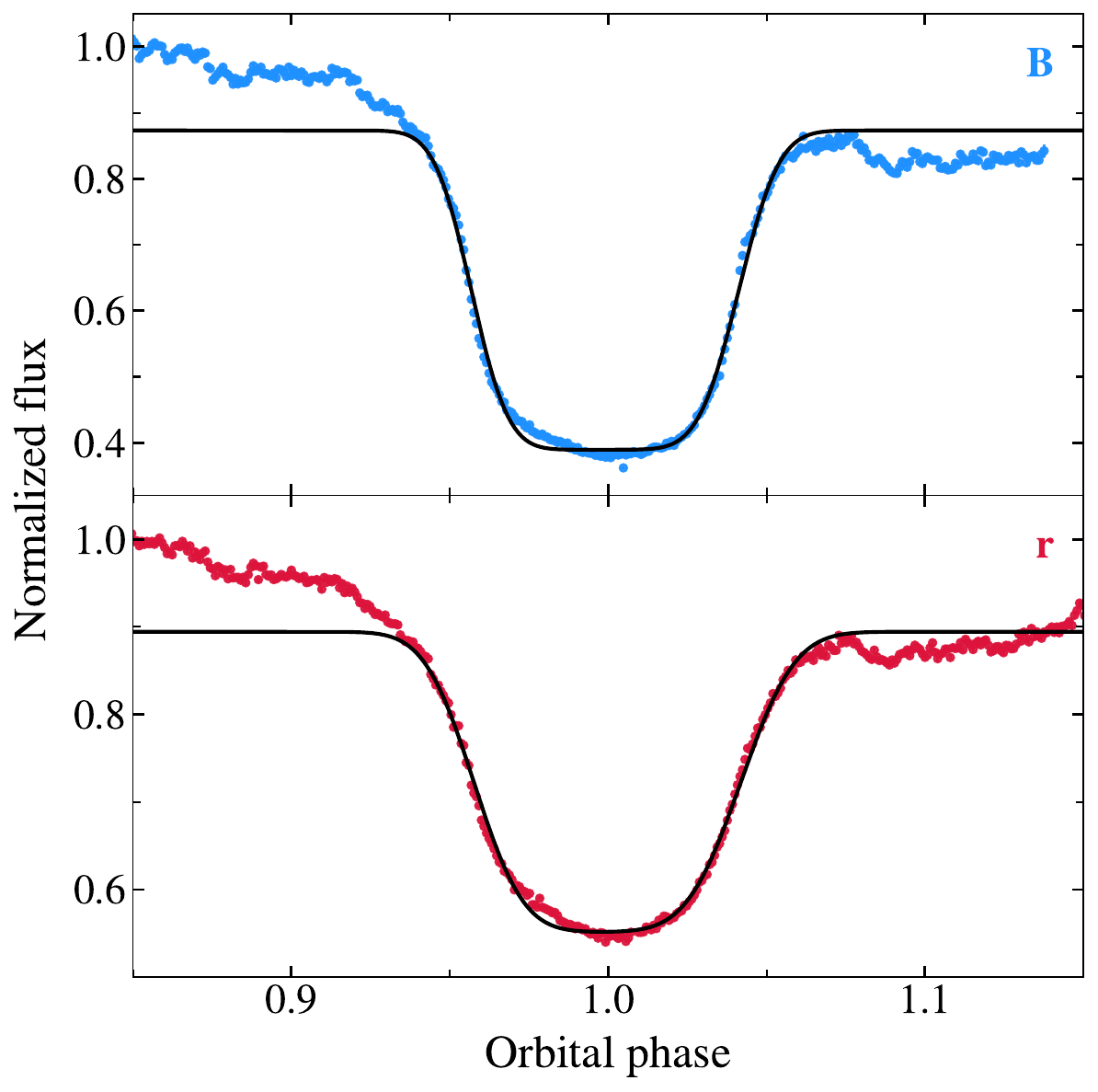}
\caption{Orbital phase-folded light curves of J1152 obtained from B-, and r-band data obtained on 14 June 2025. The solid black lines represent the best-fit model to each band.}
\label{fig:folded-saao}
\end{figure}

\par In addition, we divided the \textit{TESS} light curve into one-day segments in order to examine the detailed evolution of the eclipse profiles. Given the cadence and signal-to-noise of the \textit{TESS} data, a Gaussian profile provides an adequate representation of the eclipse shape, whereas higher time-resolution data require more detailed models. Each one-day phase-folded and binned profile was fitted with a Gaussian model, which enabled us to determine the fractional eclipse depth ($\delta$), mid-eclipse phase ($\phi_{0}$), and eclipse phase width at half depth ($\Delta \phi_{1/2}$) on a daily basis. The temporal evolution of these parameters is presented in Fig. \ref{fig:eclipse}. We find that during S1, the mid-eclipse phase occurs slightly later than phase $\sim$1.0, with an average value of $1.006 \pm 0.002$, whereas during S3, it occurs slightly earlier, with an average of $0.9945 \pm 0.0005$. The weighted mean of $\phi_{0}$ across all segments is 0.9953 $\pm$ 0.0004, indicating a small offset from phase 1.0.  The eclipse phase width decreases as the system evolves from S1 to S2 and remains roughly constant from S2 through S4. The average phase widths during S1 and S3 are $0.146 \pm 0.006$ and $0.094 \pm 0.001$, respectively. Moreover, the eclipse depth increases from S1 to S2, reaches a maximum at S3, and decreases slightly to a value comparable to S2 during S4. The average fractional eclipse depths for S1, S2, S3, and S4 are $0.0273 \pm 0.0008$, $0.082 \pm 0.001$, $0.104 \pm 0.001$, and $0.080 \pm 0.003$, respectively.

\par The orbital phase–folded light curves of J1152 in the B- and r-bands are shown in Fig. \ref{fig:folded-saao}. Both bands display a well-defined, symmetric eclipse centered at phase 1.0. In contrast to the Gaussian profiles that adequately reproduce the broader, lower-cadence TESS data, the rounded trapezoidal model provides a significantly better fit to the higher time-resolution SAAO light curves, capturing both the flat-bottomed minimum and the smooth ingress and egress transitions. A visual inspection of the light curves indicates an average eclipse duration of approximately 1.2 h.  The eclipse depth is strongly wavelength dependent, with the B-band showing a deeper minimum ($\delta$ = 0.554 $\pm$ 0.001) than the r-band ($\delta$ = 0.384 $\pm$ 0.001). The corresponding eclipse phase widths at half depth, derived from the rounded trapezoidal fits, are 0.08445 $\pm$ 0.00004 (B) and 0.08431 $\pm$ 0.00006 (r). The deeper and slightly sharper B-band eclipse indicates that the occulted region is dominated by a hotter, compact source such as the WD, hot spot, and the innermost accretion disc, while the shallower r-band profile traces additional contribution from cooler, more extended disc regions. The slanted features during eclipse minima in both light curves resemble the asymmetries expected from a hot-spot contribution, similar to those reported by \citet{littlefair2007,littlefair2008} in other eclipsing DNe.




\subsection{Average spectroscopy} \label{sec3.3}
\begin{figure}
\centering
\includegraphics[width=0.5\textwidth,height=0.4\textwidth]{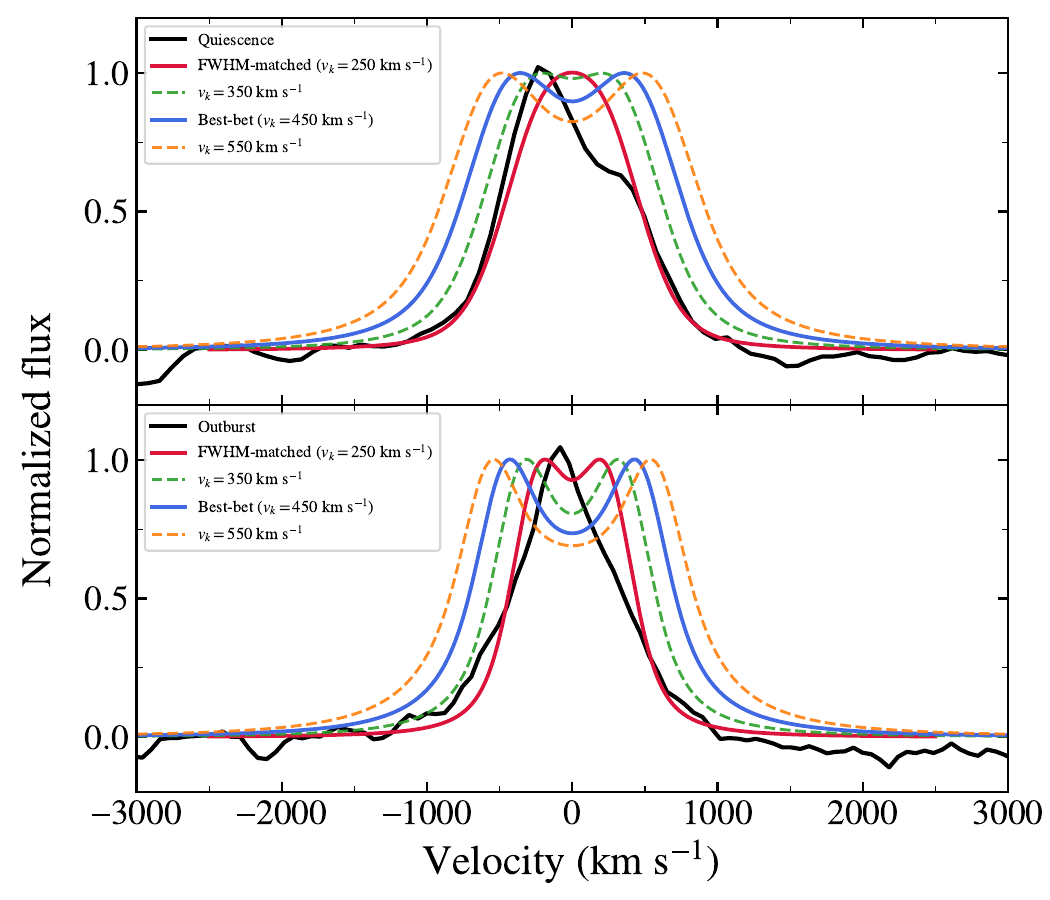}
\caption{H$\alpha$ line profiles in quiescence and outburst, compared with illustrative accretion-disc models convolved to the instrumental resolution for different v$_{\rm k}$.}
\label{fig:line_profiles}
\end{figure}

The average optical spectrum obtained during the outburst is shown in Fig. \ref{fig:spectra}, together with the quiescent spectrum for comparison. The optical spectra of J1152 exhibit strong, single-peaked Balmer emission lines from H$\alpha$ to H$\gamma$ during outburst. In quiescence, the Balmer emission lines do not appear to exhibit a clear double-peaked structure. To assess whether these emission lines are consistent with double-peaked accretion-disc profiles (see Fig. \ref{fig:line_profiles}), we compared the observed line shapes with illustrative optically thin disc models. The models assume an emissivity law proportional to r$^{-1}$ and are convolved to the instrumental resolution. Although this emissivity law is simplistic, we note that varying this prescription does not resolve the difficulty in reproducing the observed line profiles. The outer-disc Keplerian velocities (v$_{\rm k}$) were drawn from the range implied by the mass ratio and inclination constraints derived in Section \ref{sec3.6}; a representative `best-bet' value, corresponding to the average of this range, is highlighted in Fig. \ref{fig:line_profiles} for reference.  The quiescent line profiles do not show unambiguous double-peaked structure, but remain broadly consistent with weak or asymmetric double-peaked emission once resolution effects are taken into account. In contrast, the outburst profiles are predominantly single-peaked and are not well reproduced by any reasonable double-peaked disc model at the available resolution. This behaviour is most pronounced in H$\alpha$, which is likely affected by additional low-velocity emission that fills in the line core. We have also examined the higher-order Balmer lines (H$\beta$ and H$\gamma$), which show somewhat more disc-like structure, with marginal indications of shoulders in outburst. However, the peaks are not clearly resolved in these lines, and they remain inconsistent with the well-separated double-peaked morphology predicted by the disc models. In addition to the Balmer lines, the neutral helium lines He \rn{1} 5875 \AA\ and He \rn{1} 6678 \AA\ are present during quiescence. Notably, the He \rn{2} 4686 \AA\ line is significantly enhanced during outburst, while it is only marginally detected in quiescence. We determined the identification, flux, equivalent width (EW), and FWHM of the principal emission lines using single-Gaussian fits; these measurements are listed in Table \ref{tab:spectra}. The uncertainties associated with each parameter represent the standard deviations derived from multiple spectral measurements. 


\subsection{Radial velocity fits } \label{sec3.4}
\begin{figure}
\centering
\includegraphics[width=0.5\textwidth]{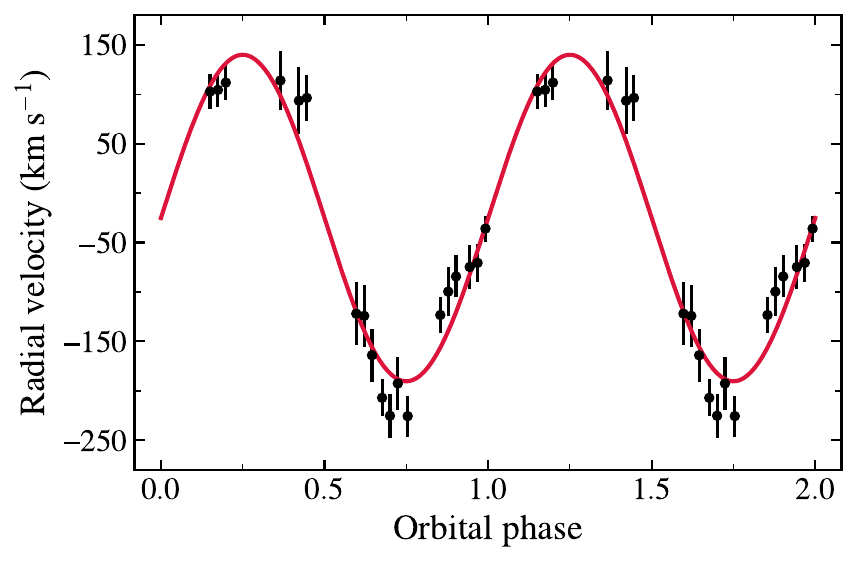}
\caption{Absorption-line radial velocities folded on the eclipse ephemeris, with the best-fitting sinusoid superposed.}
\label{fig:velocity_fit}
\end{figure}

We measured radial velocities of several emission lines using convolution line-centering algorithms \citep{schneider1980}, but did not obtain useful results.  However, the spectrum shows absorption features of a late-type star, around type K. Using the task {\tt fxcor} in the {\tt IRAF} package {\tt rv}, we measured velocities of the late-type component by cross-correlating our spectra with an average of K-type velocity standards that had been shifted to apparent zero velocity before combining. The wavelength region correlated was from 5000 to 6300 \AA , omitting a 100 \AA\ region around the strong NaD lines to avoid possible contamination from He \rn{1} 5875 \AA ~emission.  Nineteen of our 23 spectra yielded useful absorption velocities. All times are in BJD and velocities are barycentric-corrected. We fitted these velocities with a sinusoid of the form 
$$v(t) = \gamma + K \sin[2 \pi (t - T_0) / P],$$
with $P$ fixed at the eclipse period (equation ~\ref{eqn}).  The 
best-fit parameters were $\gamma$ =$-25\pm5$ km s$^{-1}$,
$K$ =$165\pm8$ km s$^{-1}$, and $T_0$ =${\rm BJD } ~2460730.235\pm0.003$.  Fig.~\ref{fig:velocity_fit} shows the absorption-line velocities and the best fit. The radial velocity derived $T_0$ is consistent with the eclipse ephemeris (equation ~\ref{eqn}), with a small phase offset of $\Delta \phi$=$-0.018\pm0.007$, indicating no significant difference between the spectroscopic and photometric ephemerides.

\par
\begin{figure}
\centering
\includegraphics[width=0.5\textwidth,height=0.4\textwidth]{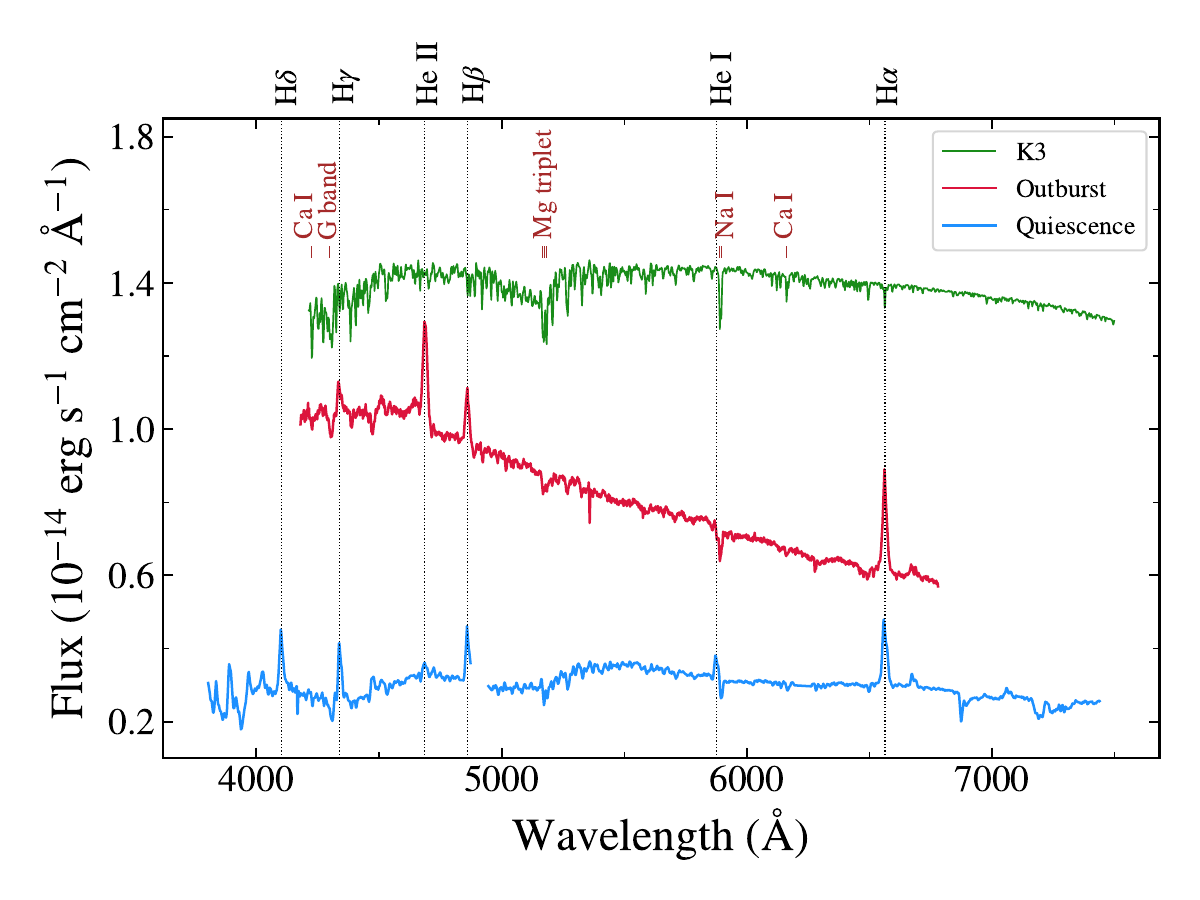}
\caption{Comparison of the optical spectra of J1152 obtained during the quiescent and outburst phases. The outburst spectrum is the average spectrum shown in the donor-star rest frame. A K3 template spectrum is overplotted and offset vertically for clarity. } 
\label{fig:spectra}
\end{figure}

 The average outburst spectrum in Fig. \ref{fig:spectra} shows weak features of the secondary star.  To quantify the secondary's contribution and constrain its spectral type, we began by shifting the individual flux-calibrated spectra to their rest frames, using the absorption-line velocity fit parameters found in earlier analysis and averaging the shifted spectra.  We then subtracted away existing spectra of K-type stars classified by \citet{keenan1989}, systematically varying the scaling factor and spectral type, and finally examined the results interactively with the aim of optimising the cancellation of the secondary features.  The strong contribution from the outbursting disc made this difficult to judge precisely. Acceptable results were found around K3$\pm$2 subtypes,
with the secondary contributing 25 to 35 per cent of the flux at 6500 \AA.  The total flux was $f_{\lambda}(6500)$ =6.1$\times10^{-15}$ erg cm$^{-2}$ s$^{-1}$ \AA $^{-1}$. Converting this to an Oke AB magnitude and equating it to SDSS r \citep[see][]{fukugita1996} gives r =14.05 mag.  Taking the secondary's contribution to be 30 per cent implies r $\sim$16.3 mag for the secondary alone.  This is a rather rough estimate, mostly because of the uncertain decomposition. 




\begin{table*}
\centering
\caption{Identification, flux, EW, and FWHM for emission features in the spectra of quiescent and outburst phases.}
\label{tab:spectra}
\renewcommand{\arraystretch}{1.4}
\begin{tabular}{lcccccccccc}
\hline
\multirow{2}{*}{Identification} && \multicolumn{3}{c}{Quiescent} && \multicolumn{3}{c}{Outburst$^{\dagger}$} \\
\cline{3-5} \cline{7-9}
 && Flux & -EW & FWHM && Flux & -EW & FWHM \\
 \hline
H $\delta$ (4102 \AA) && 3.761 $\pm$ 0.009 & 13.97 $\pm$ 0.04 & 1593 $\pm$ 9 &&  ... & ... & ... \\ 
H $\gamma$ (4340 \AA) && 2.585 $\pm$ 0.006 & 10.12 $\pm$ 0.02 & 1103 $\pm$ 3 &&  1.556 $\pm$ 0.078  & 1.51 $\pm$ 0.08 & 1068 $\pm$ 68\\
He \Romannum{2} (4686 \AA) && p & p & p && 8.243 $\pm$ 0.049 & 8.46 $\pm$ 0.06 & 1603 $\pm$ 10\\
H $\beta$ (4861 \AA) && 3.098 $\pm$ 0.068 & 10.43 $\pm$ 0.22  & 1225 $\pm$ 49 && 4.125 $\pm$ 0.051 & 4.45 $\pm$ 0.06 & 1421 $\pm$ 17\\
He \Romannum{1} (5875 \AA) &&  1.195 $\pm$ 0.017 & 3.84 $\pm$ 0.06  & 859 $\pm$ 17 && p & p & p \\
H $\alpha$ (6563 \AA) && 4.572 $\pm$ 0.043 & 15.91 $\pm$ 0.20  & 1137 $\pm$ 7 &&  6.016 $\pm$ 0.049 & 10.11 $\pm$ 0.10 & 995 $\pm$ 7 \\
He \Romannum{1} (6678 \AA) && 0.663 $\pm$ 0.024 & 2.29 $\pm$ 0.08 & 734 $\pm$ 37 &&  ... & ... & ... \\
\hline
\end{tabular}

\bigskip
\emph{Note. 1.} Flux, EW, and FWHM are in the unit of $10^{-14}$ $\rm erg ~cm^{-2}$ $\rm s^{-1}$, \AA, and $\rm km ~s^{-1}$, respectively.\\
\emph{2.} `p' -- line present but S/N too low for measurement.\\
\emph{3.} $\dagger$ denotes parameters derived from the combined spectrum obtained during the outburst.

\end{table*}

\subsection{Phase-resolved Spectroscopy} \label{sec3.5}
\begin{figure*}
\centering
\includegraphics[width=0.8\textwidth]{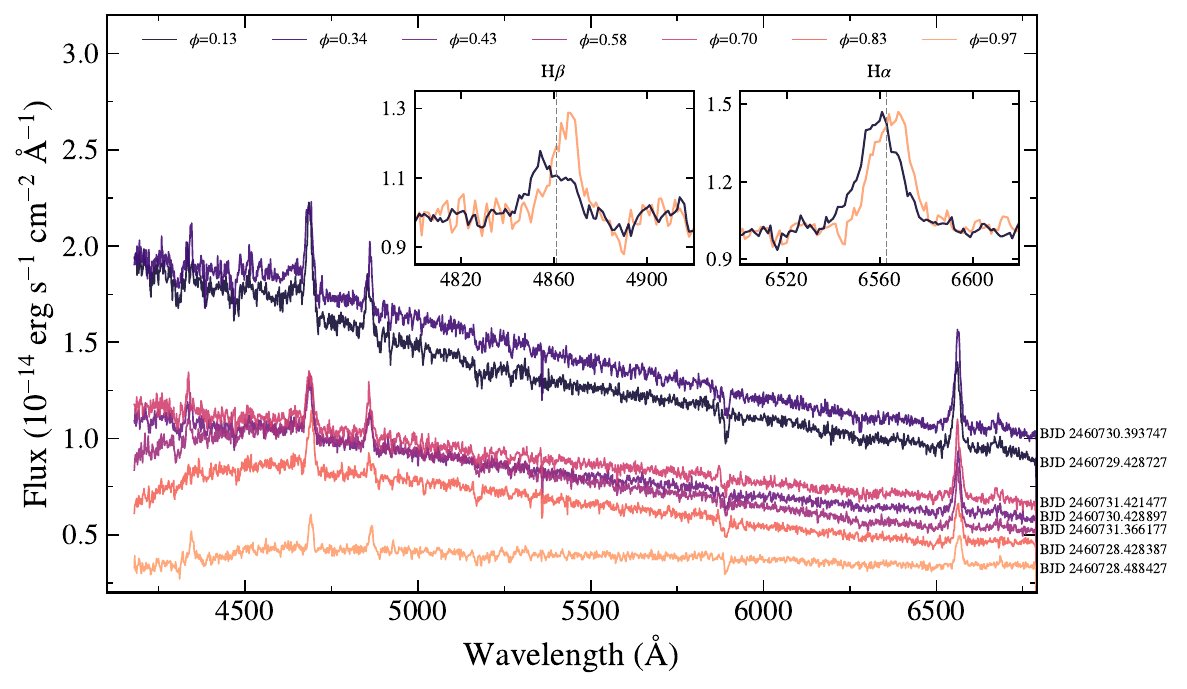}
\caption{Phase-resolved optical spectra obtained during outburst, showing the evolution of the continuum and emission lines across the orbit. The H$\alpha$ and H$\beta$ line regions are shown in the inset panels. The inset spectra are normalized to highlight changes in the line profile morphology between pre- and post-eclipse phases.}
\label{fig:phase_resolved_spectra}
\end{figure*}

\begin{figure}
\centering
\includegraphics[width=0.5\textwidth]{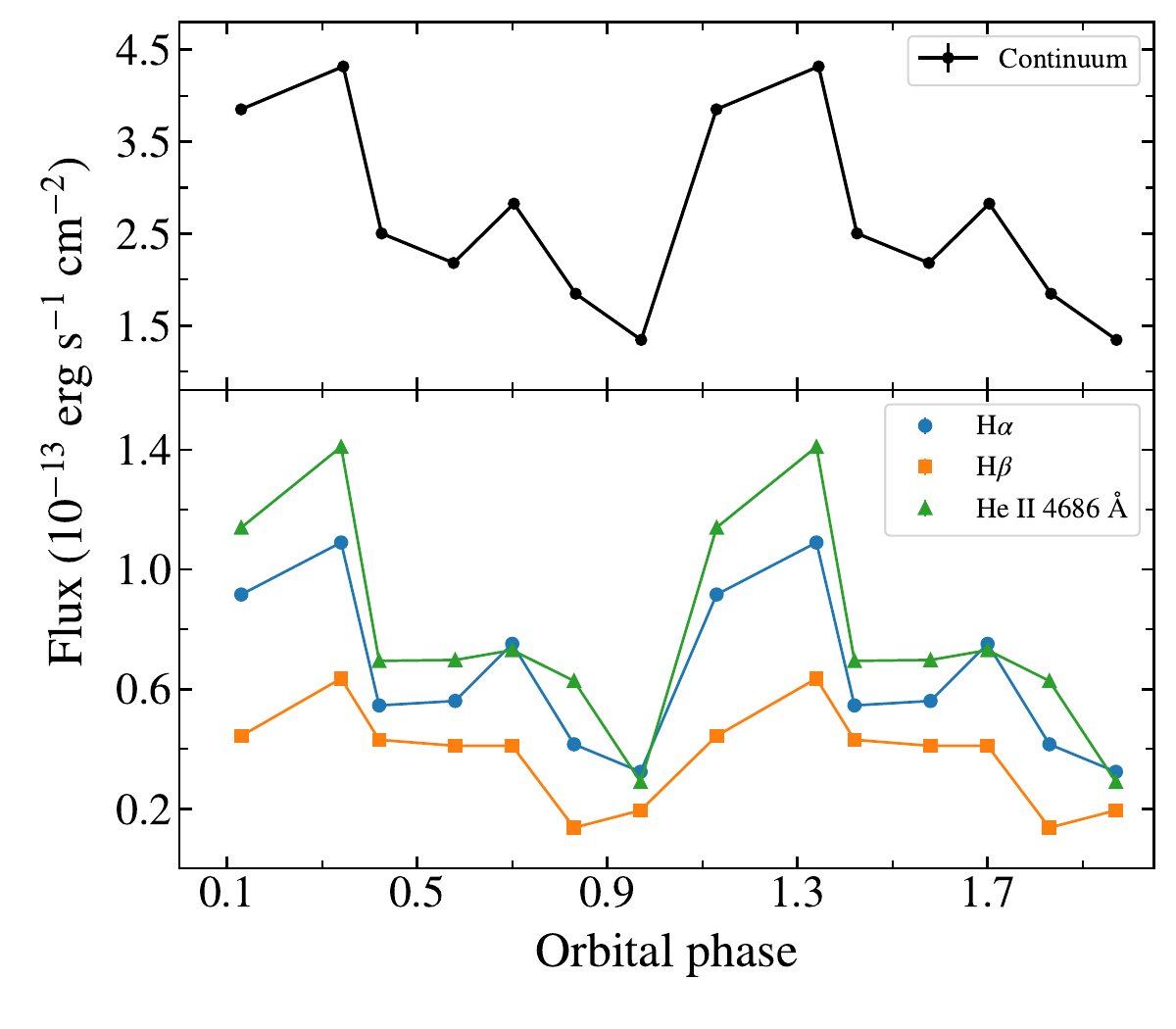}
\caption{Flux variations of continuum, H$\alpha$, H$\beta$, and He II 4686 \AA ~as a function of orbital phase during outburst.}
\label{fig:phase_line_flux}
\end{figure}

Fig. \ref{fig:phase_resolved_spectra} shows `representative' phase-resolved optical spectra obtained during outburst, sampling seven orbital phases between $\phi$ = 0.13 and 0.97. Clear variations are seen in both the continuum and the emission lines as the system approaches eclipse. The continuum level decreases steadily with phase, reflecting the occultation of the accretion disc by the secondary. The Balmer and He\,\rn{2}~4686\,\AA\ emission lines exhibit similar behaviour, showing a marked reduction in flux near $\phi$$\approx$1.0 (Figs.  \ref{fig:phase_resolved_spectra}  and \ref{fig:phase_line_flux}). Despite this decline, the emission lines do not disappear entirely during the eclipse. Even at $\phi$=0.97, strong H$\alpha$, H$\beta$, and He\,\rn{2}\,4686 \AA ~emission remains visible, immediately indicating that the line-forming regions cannot be confined to the innermost parts of the accretion disc. A closer comparison of the H$\alpha$ and H$\beta$ line profiles before ($\phi$=0.97) and after eclipse ($\phi$=0.13) suggests weak, phase-dependent differences in the line wings on either side of eclipse, consistent with a possible rotational disturbance (see, inset panels of Fig. \ref{fig:phase_resolved_spectra}). We emphasize that this indication is subtle and would require higher time-resolution and higher S/N data to confirm unambiguously. 
\par We also derived the line fluxes for all three emission lines, together with the optical continuum, across the orbit, as shown in Fig. ~\ref{fig:phase_line_flux}. To quantify the extent of uneclipsed emission, we estimated the uneclipsed fraction of the optical continuum, measured in a line-free region around 6500\,\AA, and of the continuum-subtracted emission lines by taking the ratio of the mid-eclipse flux (averaged over orbital phases 0.9--1.1) to the out-of-eclipse flux (averaged over phases 0.2--0.4 and 0.6--0.9). The continuum near 6500\,\AA\ exhibits an eclipse depth of approximately 50 per cent, indicating that roughly half of the continuum emission remains visible at mid-eclipse. The emission lines show uneclipsed fractions of $\sim$38--57 per cent, demonstrating that a substantial fraction of the line-emitting gas arises from regions not occulted during eclipse.

\subsection{Constraints on the system geometry} \label{sec3.6}
\begin{figure}
\centering
\includegraphics[width=0.5\textwidth, height=8cm]{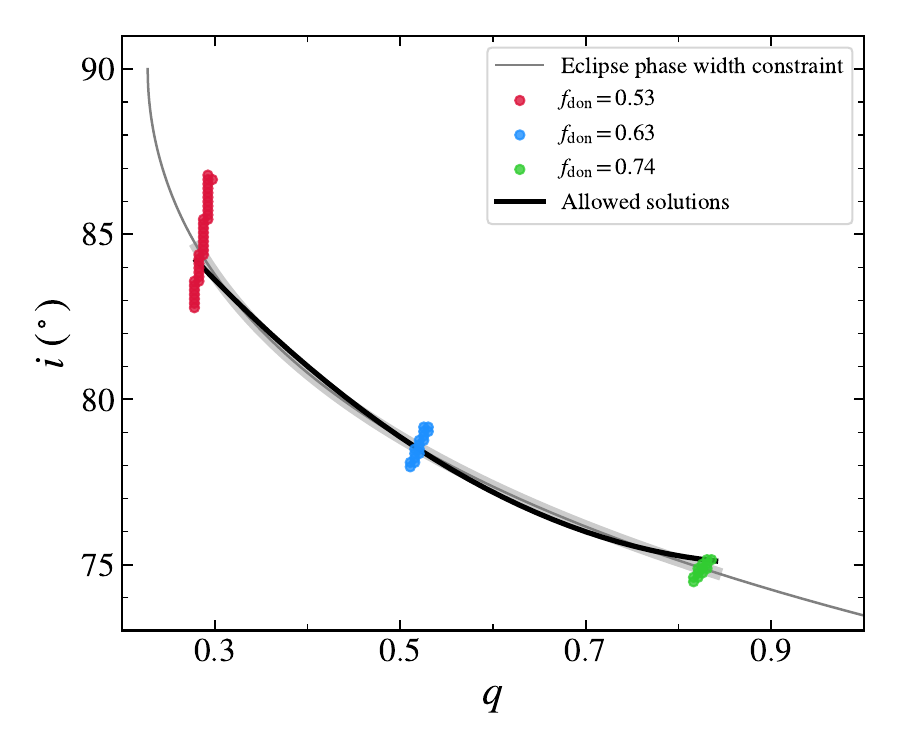}
\caption{Constraints on the system geometry in the $q$--$i$ plane. The thin grey curve shows the eclipse phase-width constraint, coloured points mark the joint solutions for different donor fractions, the thick grey line indicates the allowed family of solutions, and the thick black curve shows a quadratic approximation to this family (see text for details). }
\label{fig:qi_solution}
\end{figure}

The geometry of the system is constrained by the combined requirements imposed by the eclipse phase width at half depth ($\Delta \phi_{1/2}$) and the observed ellipsoidal modulation. For a Roche-lobe-filling donor, the radius of the secondary is given by
\begin{equation}
\left(\frac{R_2}{a}\right)^2
=
\sin^2(\pi \Delta\phi_{\rm 1/2})
+
\cos^2(\pi \Delta\phi_{\rm 1/2}) \cos^2 i ,
\end{equation}
where $a$ is the orbital separation and $i$ is the orbital inclination. Assuming the secondary fills its Roche lobe, $R_2$ can be expressed as the volume-equivalent Roche-lobe radius $R_{\rm L}(q)$, given by \citet{1983ApJ...268..368E} as
\begin{equation}
\frac{R_{\rm L}}{a} =
\frac{0.49\,q^{2/3}}{0.6\,q^{2/3} + \ln\!\left(1 + q^{1/3}\right)} .
\end{equation}
Together, these equations define a continuous curve in the $q$--$i$ plane, shown as the eclipse phase-width constraint in Fig. \ref{fig:qi_solution}.
Additional restrictions arise from the ellipsoidal variability measured in the ATLAS-o band, derived using quiescent data. The ellipsoidal modulation was modelled using the full second-harmonic term of the theoretical ellipsoidal light variation, following equations~(7) and (11) of \citet{gomel2021}. In this formulation, the amplitude of the modulation depends on the mass ratio $q$, the inclination $i$, and the fractional contribution of the donor star to the total optical light, while accounting for tidal distortion, limb darkening, and gravity darkening. We adopted these coefficients appropriate for a K-type donor star. The donor contribution to the optical light is estimated to be 25--35 per cent based on the outburst spectrum (see Section~\ref{sec3.4}). After accounting for the difference in system brightness between the outburst and quiescent states, this implies a quiescent donor fraction of $f_{\rm don} \simeq 0.53$--$0.74$. For a given donor fraction, the observed ATLAS-o ellipsoidal amplitude defines a locus of allowed $(q,i)$ solutions. Combining these solutions with the eclipse phase-width constraint restricts the system geometry to a narrow segment in the $q$--$i$ plane. The resulting family of allowed solutions can be approximated by a quadratic relation,
\begin{equation}
i = 78.5 - 18.1\,(q-0.52) + 23.6\,(q-0.52)^2 ,
\label{eqn4}
\end{equation}

valid over the range $0.28 \le q \le 0.84$.


\section{Discussion} \label{sec4}
\subsection{Nature of the outburst }
The observed outburst properties indicate that J1152 is a U Gem–type DN. What makes it particularly interesting is its long orbital period ($\approx$10.46 h) and relatively small outburst amplitude ($\simeq$1.6 mag). 
The absolute magnitude of J1152 at outburst comes out to be $\simeq$4.0 mag, which is in good agreement with the values given for long orbital period DNe by \citet{Warner1987}.  The outburst light curves display highly symmetric eclipses, indicating that the optical flux during outburst is dominated by the accretion disc, with no strong contribution from azimuthally asymmetric components such as the hot spot.
DNe generally exhibit two types of outbursts -- `outside-in' and `inside-out'. The most reliable way to determine the outburst type is through the analysis of eclipse profiles, whenever such data are available \citep{menard2001}. In an `outside-in' outburst, the disc brightens first at its outer edge, and the heating wave propagates inward during the decline phase \citep[e.g.][]{vogt1983, ioannou1999}. In contrast, `inside-out' outbursts originate near the inner disc and the instability propagates both outward and inward \citep{smak1984, meyer1984, mineshige1985}. Regardless of the outburst type, the decline proceeds in a similar manner in all cases, as the cooling front originates in the outer disc and propagates inward \citep{2001cvs..book.....H}.  The outburst observed in the \textit{TESS} data of J1152 is accompanied by clear changes in eclipse behaviour, with eclipses becoming deeper and narrower during the rise and peak phases (see, Figs. ~\ref{fig:folded-tess} and~\ref{fig:eclipse}). We note that similar behaviour has been reported in systems undergoing inside-out outbursts.
We used the relation of \citet{bailey1975}, as presented in \citet{1995cvs..book.....W}, to estimate the outburst decline rate based on the system's orbital period. We also employed a more recent calibration from \citet{hypka2016}, and both relations yield decline rates of $\simeq 3.8$--$4.0~\mathrm{d~mag^{-1}}$, which are consistent with our observational data. For the \textit{TESS} outburst, we do not have a precisely measured decline duration, but we estimate a lower limit for the total outburst length of $\sim$17~days. Using the outburst-duration versus orbital-period relation from \citet{hypka2016}, originally derived by \citet{smak2000}, we obtain an expected duration of $\sim$22~days (from their equation~20), which agrees well with our lower-limit estimate and suggests that the \textit{TESS} event belongs to the category of `long' outbursts. 
\par An interesting system to consider in this context is GK Per, which is a very long orbital period ($\sim$2 d) intermediate polar, that nonetheless exhibits inside-out outbursts, despite its relatively high accretion rate. The outburst morphology is thought to result from its unusually large accretion disc \citep{kim1992}. The similarity is noteworthy because J1152 shows comparable changes during outburst, pointing to a potential role of disc size and structure in shaping the observed outburst morphology.





\subsection{Appearance of single-peaked emission lines and He \rn{2} 4686 \AA ~during outburst}
We observed clear single-peaked Balmer emission lines during outburst, while in quiescence, the Balmer lines do not show a clear double-peaked structure.  The quiescent spectrum was obtained over orbital phases $\phi$ =0.007--0.03, entirely within the eclipse window, covering $\sim$20 per cent of its duration. At these phases, the hot spot associated with the stream-disc impact region is expected to be eclipsed. However, the accretion disc is not necessarily fully obscured during eclipse, and prominent emission lines remain visible. This indicates that the line-forming region is not confined to the innermost disc regions but is instead spatially extended within the disc. The persistence of emission during eclipse suggests contributions from spatially extended regions, such as the outer disc or a disc atmosphere. A definitive discrimination between these scenarios would require detailed eclipse mapping, which we leave to future work. 

\par During outburst, if the emission lines predominantly arise in a thin Keplerian disc, their profiles should be resolvably double-peaked at inclinations as small as 15$^\circ$ \citep{horne1986a}. However, we find that even at high inclination ($i$ $\approx$75--$84^\circ$), all emission lines remain clearly single-peaked.  Interestingly, single-peaked Balmer emission lines have been seen in a special subclass of nova-like CV, known as SW Sex type stars, which are mostly eclipsing and \citet{honeycutt1986} have shown that these can originate from the wind component. Additionally, theoretical work indicates that disc winds provide a natural explanation for single-peaked optical emission lines in high-state CVs. In particular, \citet{murray1996} showed that wind kinematics can convert intrinsically double-peaked disc emission into single-peaked profiles, while radiative-transfer simulations by \citet{matthews2015} demonstrate that such profiles arise when the line-forming region is vertically extended above the disc plane.

\par Outburst spectra of DNe usually show broad Balmer absorption troughs with weak emission cores. A non-negligible subset displays He\,\rn{2}~4686\,\AA\ in outburst, at least $\sim$27\% in the sample of \citet{morales-rueda2002} and additional cases have been reported since (e.g. V364~Lib; \citealt{Kinugasa2009}; 1SWASP~J162117+441254; \citealt{Scaringi2016}; HT~Cas; \citealt{neustroev2020}; V455~And; \citealt{tampo2022}). Our outburst spectrum likewise shows strong  He\,\rn{2}~4686\,\AA\, together with a marginal Bowen blend near 4640–4650~\AA. The possible origin of He\,\rn{2}~4686\,\AA\ in the outburst has been studied by many authors and different reasons have been given in the literature. \citet{kimura2018} argued that if the WD is massive ($>$1 M$_{\odot}$), the inner disc has a relatively high temperature, due to which highly ionized emission lines are seen in some of DNe, including BV Cen \citep{wargau1988}, V364 Lib \citep{kimura2018}, and few others in \citet{morales-rueda2002}. However, in the case of a normal WD mass ($\sim$0.8 M$_{\odot}$), such lines have been observed in high inclination systems, indicating that a high WD mass is not a necessary condition. Instead, the appearance of such high-excitation lines is more naturally explained by enhanced ionization during outburst, driven by increased temperature and irradiation from the inner disc and boundary layer. In this context, \citet{morales-rueda2002} suggested the origin of He\,\rn{2}~4686\,\AA\ to be either the presence of spiral structure in their accretion discs or the irradiation of the disc’s vertically thickened sectors by the WD, boundary layer, and inner accretion disc during outburst.

\par Recently, ionization–radiative transfer simulations of disc winds by \citet{tampo2024} reproduced single-peaked He\,\rn{2}~4686\,\AA\ ~and H$\alpha$ at outburst maximum in the DN V455~And. They showed that such profiles can arise when the wind mass-loss to disc accretion-rate ratio is high ($\sim$40 per cent) and/or when the wind is clumpy with volume filling factors comparable to stellar winds. These models further imply line formation in high electron density regions. We note that the measured Balmer flux ratios are low in both states: $\mathrm{H}\alpha/\mathrm{H}\beta$ $\simeq$1.5, significantly below the Case~B low-density value of $\simeq$2.86 at $T_e$ $\sim$10$^4$\,K \citep[e.g.,][]{osterbrock2006}. Such a `flattened Balmer decrement' is naturally produced when the Balmer lines are optically thick and subject to collisional/radiative-transfer effects in high-density gas ($n_e$ $\gtrsim$10$^{11\text{–}12}\,{\rm cm^{-3}}$), conditions that are consistent with disc-wind scenarios \citep[cf.][]{tampo2022,Koljonen2023}. In our case, the observation that the emission lines are single-peaked, together with the finding that approximately one-third to one-half of the line emission in J1152 remains uneclipsed in the phase-resolved spectra, provides strong evidence for a spatially extended line-forming region rather than one confined to the inner accretion disc. This interpretation is further supported by the absence of pronounced Balmer absorption troughs during outburst. Taken together, these properties suggest that a substantial fraction of the emission arises from material located at appreciable scale heights above the orbital plane, consistent with a disc wind or vertically extended atmosphere, while not excluding a contribution from the underlying accretion disc.

\subsection{Comparison to other DNe}
CVs span a broad range of orbital periods, but the population is strongly concentrated below P$_{\rm orb}$ $\lesssim$8\,h, with a rapid decline toward longer periods \citep{knigge2011, dag2025}. Using the catalogues of \citet{hypka2016} and \citet{inight2023}, we examined the statistical distribution of known systems. When combining both datasets, only $\sim$9.5 per cent of CVs with measured orbital periods lie above 8\,h, and among these, only $\sim$4 per cent are classified as UG-type DNe. This confirms that long-period DNe represent a small and atypical subset of the overall DN population. In this context, J1152 stands out as an unusually long-period DN. Its orbital period of $\approx$10.46\,h places it among the upper few per cent of all known CVs with measured periods, and among an even smaller fraction of UG-type systems. Although the recurrence time of $\sim$40–60 days for J1152 appears relatively short for a system with such a long orbital period, it is not unusual in this regime and is consistent with the observed low outburst amplitude ($\simeq$1.6 mag). In particular, the system follows the general trend between recurrence time to outburst amplitude ratio seen in DNe. This suggests that, despite its short recurrence time, the outburst behaviour is broadly consistent with expectations. We also estimate the duty cycle using the plateau duration, obtaining a value of $\sim$0.13–0.2, consistent with the expected range for DNe with recurrence times of tens of days. We note, however, that long-period systems remain sparsely sampled, and the extent of intrinsic scatter in this regime is not yet well constrained.

\par  Only a small number of DNe have shown observational evidence for disc winds. A clear example is the eclipsing DN IP~Peg, where optical spectroscopy revealed wind components in the emission lines \citep{piche1989, marsh1990}. Disc-wind signatures have also been suggested in other long-period systems, such as OGLE-BLG504.12.201843 and V364~Lib, where single-peaked outburst spectra have been interpreted as possible evidence for a disc-wind contribution \citep{landri2022, kimura2018}. Disc-wind models for CVs typically require high disc accretion rates, of order $\dot{\rm M_{\rm d}}$ $\sim$10$^{-8}$--$10^{-7}\,\rm M_\odot\,\mathrm{yr^{-1}}$, together with wind mass-loss rates of $\dot{\rm M}_{\rm wind}$ $\sim$10$^{-9}$--$2\times10^{-8}\,\rm M_\odot\,\mathrm{yr^{-1}}$ \citep{long2002, matthews2015, tampo2024}. These values suggest that $\sim$5--10 per cent of the accreted material can be expelled through a disc wind, consistent with the expectations for high-$\dot{\rm M}$ CVs. We estimate the disc accretion rate using the inclination-corrected absolute magnitude following equation~10 of \citet{webbink1987}, obtaining $\dot{\rm M_{\rm d}}$ $\sim$$(1.7\text{--}3.6)\times10^{-8}\,\rm M_\odot\,{\rm yr^{-1}}$ for M$_{\rm V}$$\simeq$4.0 mag. This estimate assumes a secondary mass of $\sim$0.7 $\rm M_\odot$, a WD mass range of $0.83\text{--}1.28 ~\rm M_\odot$, and inclinations between $75^\circ$ and $78^\circ$ (from equation~\ref{eqn4}). This estimate is comparable to, or exceeds, the critical mass transfer rate required for thermal stability of the accretion disc  $\dot{\rm M}_{\mathrm{crit}}$ $\sim$2.4$\times10^{-8}\,\rm M_\odot\,{\rm yr^{-1}}$ (calculated using equation 13 of \citealt{dubus2018}), as expected for a disc in outburst. The inferred disc accretion rate, therefore, lies within the regime required by disc-wind models. Further, systems at P$_{\rm orb}\gtrsim$6\,h often host more massive or slightly evolved donors that drive relatively high mass-transfer rates compared to short-period CVs, although detailed evolutionary models indicate that the observed late spectral types in this regime require $\dot{\rm M_{\rm T}}$ $\lesssim$5$\times10^{-9}$\,M$_\odot\,{\rm yr^{-1}}$ \citep{baraffe2000}.  We also estimate the secular mass transfer rate using the scaling between the duty cycle and the critical accretion rate (equation~3.31 of \citealt{cannizzo1988}). The inferred values lie in the range $\dot{\rm M_{\rm T}}$ $\sim$(1.5$\times10^{-10}$–$1.6\times10^{-9}) \rm M_\odot yr^{-1}$, consistent with the expectations from  evolutionary models.

\par Compared with other well-studied long-period DNe such as AT~Ara (9.0\,h), DX~And (10.6\,h), OGLE-BLG504.12.201843 (12.5\,h), and BV~Cen (14.6\,h), J1152 sometimes shows notable similarities and sometimes differs from these systems. For instance, the secondaries in AT~Ara and OGLE-BLG504.12.201843 appear to be consistent with unevolved main-sequence stars \citep{bruch2003, landri2022}. In contrast, \citet{drew1993} found the donor in DX~And to have a K1 spectral type and a mass of $<$0.6\,M$_\odot$, which is too small to fill its Roche lobe unless the star is evolved. They inferred that the donor radius must be at least $\sim$40 per cent larger than that of a main-sequence K1 star. Similar evidence for an inflated secondary is seen in BV~Cen, where \citet{gilliland1982} estimated a donor mass of 0.9\,M$_\odot$ with a radius exceeding the main-sequence value by more than 50 per cent. Following this approach, we estimated the radius of the donor in J1152 by comparing its Roche-lobe size with that expected for a main-sequence star of similar spectral type. Assuming a representative K3 spectral type with a main-sequence mass of $\sim$0.75\,M$_\odot$, and radius of $\sim$0.8\,R$_\odot$, and adopting WD masses in the range 0.8-1.3\,M$_\odot$, we find that the donor must be $\sim$28--31 per cent larger than a corresponding main-sequence star. This inference is further supported by the mid-infrared luminosity: using the WISE W2 magnitude and a distance of 640 pc, we derive an absolute magnitude of $M_{W2} \approx$3.5, which is brighter than expected for a ZAMS K-type dwarf but significantly fainter than a subgiant or giant, consistent with a mildly inflated donor. This level of inflation strongly suggests that the secondary in J1152 is evolved, consistent with theoretical expectations for CVs in the long-period regime. Indeed, population-synthesis studies predict that roughly half of non-magnetic CVs with $P_{\rm orb} \gtrsim$6\,h host at least mildly evolved donor stars \citep{goliasch2015}. Systems with evolved donors are often associated with enhanced helium abundances, which may manifest as elevated He/H line ratios \citep[e.g.,][]{breedt2012}. In our case, the measured He \rn{1} (5875 \AA)/H$\alpha$ flux ratio ($\approx$0.26 in quiescence) lies within the typical range of 0.2--0.4 observed in DNe \citep{szkody1981} and does not indicate a clear helium enhancement. However, such line ratios do not provide a unique diagnostic of donor evolution, as they are also sensitive to the physical conditions within the accretion disc. While our measurement does not support a strongly evolved donor, it does not exclude the presence of an inflated donor. More definitive constraints would require detailed abundance studies or direct detection of donor features, which are beyond the scope of the present work given the limitations of our dataset.


\par To our knowledge, eclipsing DNe at long orbital periods are essentially unknown in the current CV population. Well-studied systems with P$_{\rm orb}$ $>$8\,h, such as AT~Ara, DX~And, BV~Cen, and OGLE-BLG504.12.201843, do not exhibit eclipses, and no confirmed eclipsing DN is presently known above this period range. J1152, therefore, appears to be the first securely identified eclipsing DN in the long-period regime. The combination of relatively deep eclipses, frequent DN outbursts, and possible indications of a disc wind contribution makes J1152 an especially compelling target for future long-term monitoring and multi-wavelength follow-up studies.


\section{Conclusions} \label{sec5}
Based on the optical timing and spectral analyses of J1152, we conclude that it is a member of the relatively rare population of long-period DNe. Our main findings are summarised below.

\begin{itemize}
    \item[1.] J1152 is a U~Gem--type dwarf nova with an orbital period of $\approx$10.46\,h, placing it in the extreme long period tail of DN population. The system exhibits recurrent, low-amplitude outbursts ($\simeq$1.6\,mag) with durations and decline rates consistent with long outbursts in U~Gem--type systems. 
    
    \item[2.] By combining the eclipse phase width at half depth with constraints from the ellipsoidal modulation, we constrain the system geometry to a narrow locus in the $(q,i)$ plane. Acceptable solutions span mass ratios $0.28 \lesssim q \lesssim 0.84$, with corresponding inclinations $i$ $\simeq$75--$84^{\circ}$.

      \item[3.] The outburst spectrum shows strong He\,\rn{2}~4686\,\AA\ emission and a flattened Balmer decrement, both consistent with dense, optically thick gas in a disc wind. These properties place J1152 among the growing number of DNe in which disc wind accretion plays a key dynamical role.

       \item[4.] Absorption features consistent with a K3-type donor star are detected, contributing $\sim$30 per cent of the flux at 6500\,\AA ~during outburst. Comparison of the donor's inferred spectral type with main-sequence expectations suggests that the secondary star is moderately inflated, consistent with a mildly evolved donor in a long orbital period system.




\end{itemize}

\section{Acknowledgements}
We thank the anonymous referee for their constructive feedback, which improved the quality of this paper. This paper includes data collected with the \textit{TESS} mission, obtained from the MAST data archive at the Space Telescope Science Institute (STScI). Funding for the \textit{TESS} mission is provided by the NASA Explorer Program. This paper uses observations made from the South African Astronomical Observatory. The 1.9-m observations were taken as part of the 2025 Dartmouth
Foreign Study Program in Astronomy; more details can be found in \citet{thorstensen2026}. Some of the observations reported in this paper were obtained with the Southern African Large Telescope (SALT). This work has made use of data from the Asteroid Terrestrial-impact Last Alert System (ATLAS) project. The Asteroid Terrestrial-impact Last Alert System (ATLAS) project is primarily funded to search for near earth asteroids through NASA grants NN12AR55G, 80NSSC18K0284, and 80NSSC18K1575; byproducts of the NEO search include images and catalogs from the survey area. This work was partially funded by Kepler/K2 grant J1944/80NSSC19K0112 and HST GO-15889, and STFC grants ST/T000198/1 and ST/S006109/1. The ATLAS science products have been made possible through the contributions of the University of Hawaii Institute for Astronomy, the Queen’s University Belfast, the Space Telescope Science Institute, the South African Astronomical Observatory, and The Millennium Institute of Astrophysics (MAS), Chile. We acknowledge ESA Gaia, DPAC and the Photometric Science Alerts Team (http://gsaweb.ast.cam.ac.uk/alerts). This work is based on the research supported in part by the National Research Foundation of South Africa. NR gratefully acknowledges support from the Southern African Large Telescope (SALT) through the SALT-Stobie fellowship for her visit to the Inter-University Centre for Astronomy and Astrophysics (IUCAA), Pune, India, as well as the local hospitality provided by IUCAA during the preparation and finalization of this manuscript. NR also thanks Arti Joshi and Alaxender Panchal for helpful discussions.  AB thanks the funding from the Anusandhan National Research Foundation (ANRF) under the Prime Minister Early Career Research Grant scheme (ANRF/ECRG/2024/000675/PMS). IM acknowledges support from RSCF grant 25-72-10176.

\section{DATA AVAILABILITY}
The \textit{TESS} data sets are publicly available in the \textit{TESS} data archive at \url{https://archive.stsci.edu/missions-and-data/tess}. The AAVSO, ASAS-SN, and CRTS data sets are available at \url{https://www.aavso.org/data-download} and  \url{https://asas-sn.osu.edu/variables}, and \url{http://nunuku.caltech.edu/cgi-bin/getcssconedbid_release2.cgi}, respectively. The optical photometric and spectroscopic data underlying this article will be shared on reasonable request to the corresponding author.

\bibliographystyle{mnras}
\bibliography{ref.bib}


\label{lastpage}
\end{document}